# EYE-TRACKING-BASED DESIGN OF MIXED REALITY LEARNING ENVIRONMENTS IN STEM

D. Sonntag, O. Bodensiek

*Technische Universität Braunschweig (GERMANY)*

## Abstract

With the advent of commercially available Mixed-Reality(MR)-headsets in recent years MR-assisted learning started to play a vital role in educational research, especially related to STEM (science, technology, engineering and mathematics) education. Along with these developments it seems viable to further frameworks and structured design processes for MR-based learning environments. Instead of a widely applicable framework for designing educational MR applications, we here consider the case of virtually enhancing physical hands-on experiments in STEM, where students are given a certain problem to solve, and how to design these. For this focused realm, we suggest an empirically driven problem- and user-centred design process for MR applications to get novices to act more like experts and exemplify it for a specific experiment and problem set containing a non-trivial electric circuit with capacitors and coils.

## 1 INTRODUCTION

As augmented reality (AR) is nowadays easily accessible via mobile devices, learning in environments based on AR are is an active field of educational research. While overall benefits of AR learning environments have been reported in many studies [1, 2], meta-analyses especially revealed learning gains by using AR compared to non-AR settings on the order of a medium sized effect (Cohen's *d=0.68* and *d=0.64*, respectively) [2, 3]. Furthermore, using AR can well foster conceptual knowledge acquisition in STEM laboratory courses [4]. For different reasons, however, most of the studies up to date have been related mobile or tablet-based AR instead of using head-mounted devices (HMD). However, only the latter enable full Mixed Reality (MR) applications that imply interaction of physical and virtual elements including interaction of users with virtual objects similar to real world interaction. Since MR-HMD became commercially available in recent years, a couple of applications and studies related to laboratory experiments in STEM education came up [5–8]. While it seems that the instructional design principles according to cognitive load theory [9] and cognitive theory of multimedia learning (CTML) [10] do also apply to MR learning environments [5], special features may arise depending on the implemented degree of physical presence and agency, which in turn mediate affective and cognitive factors as well as learning outcomes [11].

In addition, when specific problem solving is also addressed, potential problem-solving strategies are also to be considered when designing a learning environment. Problem solving is, however, highly depending on individual cognitive abilities and intrapersonal characteristics besides contextual conditions [12]. Novices and experts, for example, are likely to approach the experimental problem in distinctly different ways. Expert approaches, in turn, might yield useful hints for successful problem-solving strategies and novices could be supported by MR such that they are more likely to follow such strategies despite the lack of experience. With a view to potential future adaptive learning environments in MR, it might be beneficial to use eye-tracking functionalities of recent MR devices both for post-hoc and real-time analysis.

In this paper, we propose to extend theoretically founded design decisions of MR-enhanced hands-on experiments in STEM by empirical studies including eye-tracking - similar to user experience research. We first describe our design approach applied to a specific experiment, followed by the results of our analysis with *N=19* participants. A mixed-methods approach was conducted consisting of eye-tracking analysis, interviews and questionnaires. We then describe a generalisation of our methodology so that it can be used for the design process of other MR experimental learning environments. Our results indicate that a small to medium-sized sample of test persons may be sufficient to successfully design a meaningful MR application.

## 2 METHODOLOGY

In the according qualitative study we first carried out a conceptual knowledge test in order to divide the sample of participants (university students of STEM-related fields, *N=19*) into novices and experts. This paper-pencil test was focused on electrical circuits especially concerning the function of capacitors and coils. It also included some questions about students' mental models in the field of electricity theory. Afterwards the same students carried out an experiment while wearing an eye-tracking headset.

Figure 1 shows the experiment the participants were working on. At the bottom of the electrical circuit is a voltage source, which is preset to 12 V. In the middle branch there is a lamp and a parallel connection of capacitors. With the lower switch the lower circuit can be closed. The middle lamp glows as long as the capacitors are being charged. As soon as the capacitors are fully charged, the middle lamp turns off again. In the upper branch there is another lamp, a coil and a resistor. The upper switch is used to connect the upper branch. If the lower switch is closed and the upper switch will be closed the upper lamp will light up with a time delay. This occurs since the coil counteracts the change in current because of Faraday's law of induction. We chose this structure of the electrical circuit because we did not want neither the experts nor the novices to have worked on this or a similar circuit before. Only if we can guarantee this, we can analyse and compare their preliminary approaches to the experiment as well as to solve the given problem.

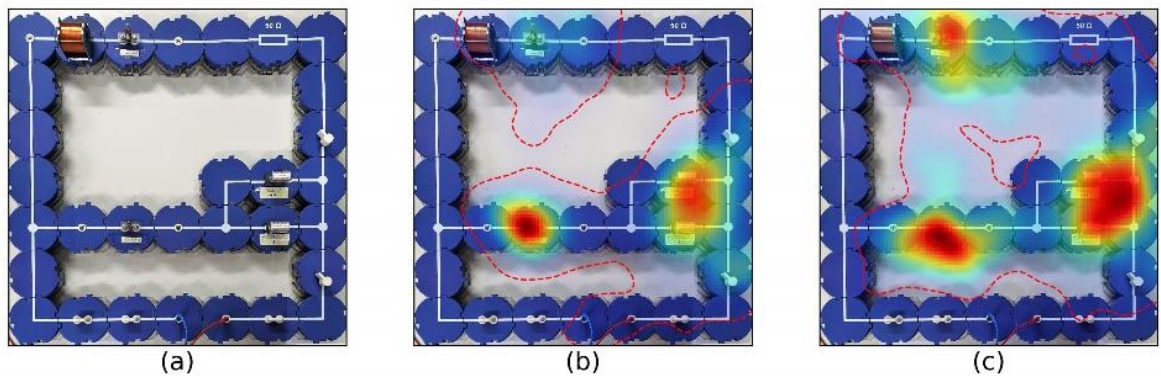

*Figure 1. (a) Electric circuit used in the study. (b) Superimposed cumulative heatmap for visual attention of novices and (c) experts. The dashed line indicates an arbitrary common threshold of gaze weight for better visual feature discrimination between novices and experts.*

While working on the electrical circuit the participants were given two tasks to solve. In the first task they were asked to think about what happens in the electrical circuit on a microscopic level. They were allowed to turn the switches and to use a voltage meter. In the second task they were given a certain problem to solve, namely to maximize the time between the middle lamp going out and the upper lamp lighting up by exploring effects of the different components and interchanging them. Therefore, they were given some additional elements, e.g. capacitors with different capacities, coils with different induction values and more resistances. After each task cognitive load was measured with an adapted variant of the cognitive load scale by Leppink et al. [13, 14].

Afterwards semi-structured interviews were conducted with the participants. They were asked to analyse their own problem-solving processes especially with respect to what they have done to understand the electrical circuit as well as if they have used any mental models to gain a better understanding. At the end they were asked which model representations and features the MR application should contain in their opinion to solve the tasks successfully.

The analysis of potentially advantageous features and functions in the planned MR application was done by comparing the different methods. First, the eye-tracking data was analysed with regard to different stages of the experimentation process. Therefore, we conducted a qualitative analysis by comparing gaze heatmaps of visual attention and learning paths as well as a quantitative analysis of the data by comparing frequencies of fixations. For the frequency of fixations, we defined different areas of interest such as upper, middle and lower circuit as well as every single component such as coil and capacitors and compared novices with experts to find differences and similarities in their approaches. Second, the semi-structured interviews were analysed. Additionally, the cognitive load with respect to problem solving was compared to the data and the individual results of the problem-solving process were taken into account. Finally, the preliminary design of the MR application was derived from these diverse data sources with the aim of supporting novices such that they are facilitated to approach the problem more in way of the experts. Some selected results of the evaluation are described in the next section.

In the further process after implementation of the MR application we carried out a minor usability-study using the thinking aloud method to improve the MR application. Therefore, we asked participants of the preliminary study to look at the experiment once more while wearing a MR-HMD. While exploring the MR learning environment they were asked to tell us what they were thinking. Based on this evaluation we were able to improve the MR application especially concerning design and computer human interaction aspects. The participants also completed a system usability questionnaire based on Brooke's system usability scale [15]. It consists of 10 items and attempts to reflect the subjective opinion of the user concerning the application. We added 5 more items about the helpfulness of the visualizations in the application concerning the understanding of the experiment and the feeling they had while working with the MR device.

## 3   ANALYSIS

Based on the conceptual knowledge test we divided the participants into 10 experts and 9 novices. We first compared the results of the scientific topics to the model representations the participants had and their cognitive load they specified after each task. By comparing the mental models of experts and novices, respectively, we found a strong significant positive correlation *(r=0.63, p=0.004)*. This matches with the expectation that experts in the field of electricity have both more and more detailed mental models than novices. We found another strong significant negative correlation between model representations and the cognitive load in the first task *(r=-0.64, p=0.003)*. We derived that those participants with a diversity of mental models can accomplish the task more easily than those with less mental models. This underlines the usefulness of including model representation in the MR application to support novices in problem solving.

By evaluating the eye-tracking data we derived the thought-related foci of novices and experts on the basis of the eye-mind hypothesis. Based on this assumption we interpreted the experimental process of both groups in comparison. Some objects were fixated in a much higher intensity by novices than by experts. The assumption that long fixated objects are those the participants think intensively about matches with the qualitative interviews afterwards where participants were asked to explain their preliminary approaches to solving the tasks. As shown in the heatmaps in Figure 1 the novices were focused mainly on the middle circuit part with the capacitors and the middle lamp while experts also focus on the upper circuit (upper lamp and coil). From the quantitative evaluation of the eye-tracking data we found a weak, however not significant, positive correlation between the focus on the upper circuit and their knowledge in electricity theory *(r=0.41, p=0.08)*. We derived that the novices were struggling with the function of the capacitors so they did not even get to the point to ask themselves about the function of the coil and the resistance in the upper circuit. That also matched with their interview statements.

The participants were additionally asked what would potentially have helped them to understand the processes in the circuit and to solve the task in the best possible way. An aspect mentioned by the majority was to visualise the electric current running through the circuit.

Finally, we deduced visualizations for the MR application taking all the above evaluations into account. Therefore, in the MR application we visualize some information about the crucial objects as the capacitors and the coil, which are important to understand in order to understand the processes in the electrical circuit. For example, for the capacitor we implemented a charge level indicator, so that the users know at any time what the current charge level is. That will help them to understand why the middle lamp is turning off shortly after closing the lower switch. For this purpose, the MR application is retrieving live measurement data of the circuit.

Furthermore, we concluded from our data that it would be helpful to visualise the electric current only on a low level with electrons running through the wire conductors in the electric circuit as shown in Figure 2. With this visualisation we know at any time where electricity flows and how much. In addition, we visualize line graphs of the voltage values over the capacitors and the coil. As conclusion of our study there are a lot more features we optionally visualize in the application e.g. the magnetic field of the coil or the potential curve along the wire conductors over the whole electric circuit.

Every user is able to select and deselect individually in a menu which visualisation he or she wants to look at. There is a couple of visualisations selected from the start which the user can adapt to his level of understanding during the experimental process. Thereby we want to avoid cognitive overload of the users.

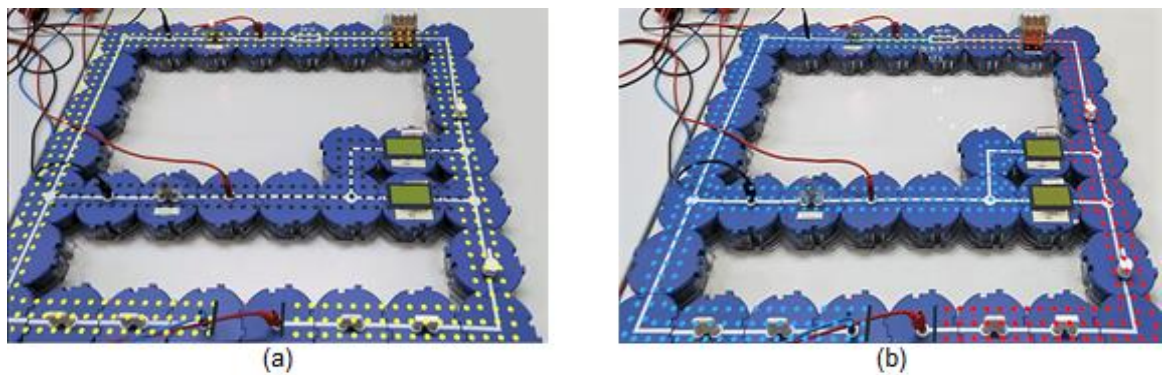

*Figure 2. MR learning environment showing (a) the electrical current, the fully loaded capacitors and (b) additionally the potential along the wire conductor.*

With the subsequent usability study, we were able to improve some design aspects of the MR application such as the structure of the menu so that every user can easily control the application. In terms of the system usability scale we had an average score of 83.60 % while the lowest value was 78 %. This corresponds to a good rating already for the first version of the MR application and we expect it to be even higher after the implementation of the usability improvements. Furthermore, all of the participants reported that the visualizations in the MR application are helpful to understand the experiment. Another important aspect is, that the whole group enjoyed working in the MR learning environment even though the technology was completely new to them.

## 4 CONCLUSIONS

Based on the previous sections we propose the following generalized design process to design MR learning environments in STEM for experiments to support novices are facilitated to solve the problem more like experts. The design process is summarized in Figure 3. The experiment must be conducted by both experts and novices to analyze and compare their approaches and to find similarities and differences.

In a first step the participants have to be divided into novices and experts based on a conceptual knowledge (paper-pencil) test. This test should include questions about the domain and problem specific topics as well as knowledge about suitable model representations. The division of our sample is important when it comes to the evaluation where the approaches of novices are compared to those of experts.

The second step is to let the students work on the experiment in general. For an easier evaluation we propose to capture two phases of the experimentation process. The first phase is the beginning when the student is trying to get an overview of the experiment. This phase is of high importance since novices do usually not manage to grasp the whole experiment and they get stuck on certain single elements. In exactly such situations an MR application can help by visualising models of important and complex elements or concepts. The second phase is about the problem-solving process where students can modify the experiment to get a solution. The aim of recording the eye-tracking data during the experiment is to trace the learning path and areas of interest of both experts and novices and to find differences.

After the experimentation phase a semi-structured interview is conducted about the preliminary approaches of the students and the model representations they used during the problem-solving process. Additionally, the students are encouraged to mention ideas and wishes for an improved MR application.

In the fourth step the evaluation of the raised conceptual knowledge, model representations, eye-tracking data and interviews is conducted. In this most important step all the empirical data merge in a joint evaluation. Based on the eye-tracking data, areas of interest are created and the corresponding weights of visual attention (gaze) are compared. To support the quantitative analysis visually, heatmaps of gaze are generated. Additionally, learning paths can be analyzed on the basis of eye-tracking data. This step creates the foundation for the design process of the MR application.

The design of the MR application takes part in step 5 and is completely based on step 4. The better the analysis is performed in step 4, the better and more profound the design process can be. In this step it is also important to consider design principles of CTML in order to avoid cognitive overload, e.g., avoid having too many visualizations at the same time.

After the design of the MR application its implementation follows. In the subsequent usability survey the focus is set more to computer-human interaction aspects than to domain- and problem-specific learning since the according visualizations have already been decided in step 4 and 5. It is recommended to use the thinking aloud method to get a direct and better feedback with only a small sample. However, a system usability scale can be used to get a quantitative statement of the usability friendliness.

In the last step of our design process the MR application is revised with the focus set to the results from the previous step to enhance the usability. Step 7 and 8 can be repeated until no further improvement is needed.

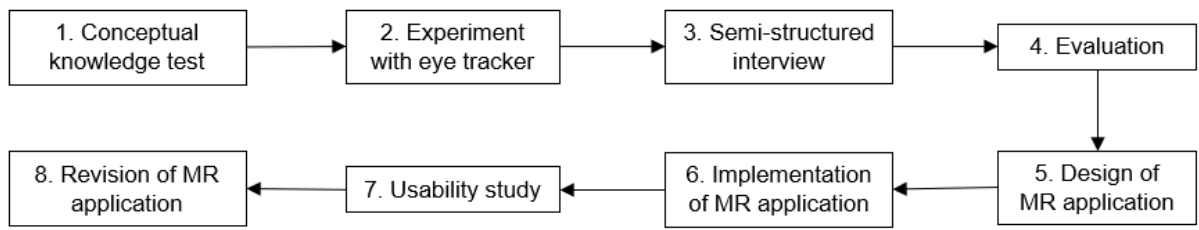

*Figure 3. User-centred and empirically driven design process of MR learning environments.*

## 5   SUMMARY

Although research shows that MR learning environments have great potential to support learners in laboratory experiments, it still lacks frameworks and design processes in this field. Therefore, in this paper we proposed a new framework for the design of corresponding MR applications. Our framework consists of eight subsequent steps using eye-tracking data and interviews to trace the experimental problem-solving approach of novices compared to experts. Based on these data MR applications can be designed to support novices in problem-solving. A final usability study can help to enhance the MR application and reduce the overall cognitive load of the user while working in the MR environment.

We furthermore presented a case study using this framework for designing an MR learning environment for a specific experiment in the realm of electricity. All of the participants of our usability study were convinced of the usefulness of the MR application to enhance the understanding of the learners while experimenting. In our study we were able to show, that our framework can lead to well elaborated MR applications. The MR application developed in this framework is currently evaluated more extensively in a follow-up study.